\documentstyle[12pt] {article}
\begin {document}
\parindent=15pt
\begin{flushright}
IPPP/02/15\\
DCPT/02/30 \\
12 March 2002
\end{flushright}
\begin{center}
\vskip 1.5 truecm
{\bf SECONDARY REGGEONS IN PERTURBATIVE QCD}\\
\vspace{.5cm}
  M.G.Ryskin and  A.G.Shuvaev \\
\vspace{.5cm}
Department of Physics and Institute for Particle Physics Phenomenology,\\
University of Durham, Durham,  DH1 3LE\\
and\\
Petersburg Nuclear Physics Institute, \\
Gatchina, St.Petersburg 188300 Russia \\
\end{center}
\vspace{1cm}
\begin{abstract}
Using the Kirschner-Lipatov equation as a basis, we study high energy elastic
quark-quark amplitudes, which correspond to secondary Reggeons and pion
exchange in the double-logarithmic approximation of perturbative QCD.
\end{abstract}
1. The high energy behaviour of elastic scattering amplitudes is described
usually in terms of  $t$-channel Reggeon pole (and cut) exchange,
that is in terms of the singularities in the complex moment $j$-plane
(see, for example, \cite{1}). The rightmost singularity, the Pomeron, has
the  quantum numbers of the vacuum and is placed at $j \approx 1$. In the framework
of perturbative QCD the corresponding singularity is formed by the two
(reggeized) gluon cut and it's properties are given by the
BFKL equation \cite{2,3} (see \cite{4,5} for more details).

The secondary Reggeon contribution corresponds to an amplitude with
quark-antiquark pair $t$-channel exchange. The leading double-logarithmic
(DL) asymptotics of this amplitude was calculated in perturbative QCD
by Kirschner-Lipatov \cite{6,7}. It appears that the quark-antiquark cut
occurs at
 \begin{equation}
 j = \omega_0(t) = \sqrt{2C_F \alpha_S/\pi}
 \end{equation}
 for the positive
signature amplitude. Here $\alpha_S$ is the QCD coupling,
$C_F = (N_c^2-1)/2N_c$ and $N_c = 3$ is the number of colors.

In the $N_c \to \infty$ limit, the positive (even) and negative (odd) signature
DL amplitudes are degenerate and both are given by the sum
of the ladder-type Feynman diagrams. For  finite $N_c = 3$, there appear
a non-ladder DL contribution to the odd signature amplitude, which
shifts the position of the singularity in $j$-plane. However this
non-ladder DL contribution is not large; numerically
$\omega_0^{(-)}\simeq 1.04\, \omega_0^{(+)}$, while for a reasonable value
of $\alpha_S = 0.25$ the effective intercept $\omega_0^{(+)} \simeq 0.46$.
This is close to the phenomenological intercepts of the $\omega$,
$\rho$ (odd signature) and $A_2$ (even signature) trajectories --
$\alpha_R(0) \approx 0.5$ which are approximately degenerate. Besides
 the $\omega$, $\rho$, $f$ and $A_2$ - Regge poles there are the contributions
with other $t$-channel quantum numbers, corresponding to $\pi$ or
$A_1$ ( unnatural $P$-parity) exchange. In the
present paper we discuss the properties of such contributions in the
framework of the DL approximation of perturbative QCD.

\noindent
2. Recall that the DL Kirschner-Lipatov quark-quark amplitude
\cite{6} may be written as the Born amplitude $A_B$ (shown by continuous
lines in Fig.1) multiplied by a DL function
$F_{DL}(\alpha_S \ln^2 s/|t|)$ which collects the loop contributions
caused by the additional gluons. For example, for the simplest colour singlet
(in $t$-channel) amplitude, the function $F_{DL}$ reads
\begin{equation}
\label{2}
F_{DL}^+\,=\,\frac{I_1(2n)}{n},\qquad
n\,=\,\sqrt{\frac{C_F\alpha_S(t)}{2\pi}}\,\ln\frac{s}{|\,t\,|}\, .
\end{equation}
Thus

\begin{equation}
\label{1}
A^+(s,t)\,=\,A_B\,\frac{I_1(2n)}{n},
\end{equation}
where
$$
A_B \,=\, -\frac 49\cdot 4\pi\alpha_S.
$$
The factor 4/9 comes from averaging over the colours of the initial quarks.
Since  the Bessel function $I_1(2n) \propto e^{2n}$ at large $n$, the high
energy ($\sqrt s$) asymptotics of the amplitude (\ref{1},\ref{2}) takes the form
$A^+ \propto s^{2n} = s^{\omega_0^{(+)}}$.

Note that to DL accuracy, the additional gluons (shown in Fig.1b
by the dashed lines) are emitted just by the colour charge of the quark
and do not change either the spin structure of the original Born amplitude
nor the isospin of the $t$-channel $q\overline q$-pair. Therefore, to select
the contribution with  known quantum numbers in $t$-channel, it is enough
to use the Firz relation
\begin{equation}
\label{3}
\gamma_\mu\,\times \,\gamma_\mu\,\biggl|_{\,s-{\rm channel}}\,\to\,
1\,\times\,1\,-\,\frac 12\,\gamma_\nu\,\times \,\gamma_\nu\,+\,
\frac 12\,i\gamma_\nu\gamma^5\,\times \,i\gamma_\nu\gamma^5\,-\,
\gamma^5\,\times \,\gamma^5\,\biggl|_{\,t-{\rm channel}} \, .
\end{equation}
As is seen from (\ref{3}), there is no tensor ($\sigma_{\nu\nu^\prime}$)
component in the $t$-channel expansion of the Born amplitude. From the
phenomenological viewpoint we have to consider the $A_2$-trajectory as the
even signature component of amplitude generated by the vector ($\gamma_\nu$)
$t$-channel current. The $\rho$ and $\omega$-trajectories originate
from the same vector ($\gamma_\nu$) current but with odd signature.

At the Born level,  relation (\ref{3}) means that the amplitudes which
 correspond to  natural and unnatural $t$-channel parities (that is
 $\gamma_\nu$ and   $\gamma_\nu \gamma_5$ or $1$ and $\gamma_5$ terms in (4)
) are degenerate.  Of course, besides this amplitude, we have the
interaction of quarks mediated by $t$-channel two (and more) gluon exchange,
that is by  Pomeron (and Pomeron cuts). Moreover the degeneracy between
the vector and axial (corresponding to the $\gamma_\nu \gamma_5$ term in
(\ref{3})) amplitudes may be spoiled by  higher order loop corrections
coming from the region of rather large distances (small quark transverse
momenta $k_t \le 1$~GeV), where the effects of confinement are not negligible.
This point was discussed in \cite{8}. Contrary to the pure "soft" Regge
phenomenological approach (where intercept of the $A_1$-trajectory
$\alpha_{A_1}(0) \le 0$ \cite{9} is lower than that for the $\rho$-trajectory
$\alpha_\rho(0) \simeq 0.5$) at  large scales, the effective (isovector)
intercept in the axial ($\gamma_\nu \gamma_5$) channel is expected to be
a little larger than that in the vector channel \cite{8}, due to the $s$-channel pion
contribution coming from the boundary between the perturbative and $k_t \le 1$~GeV
confinement region .

\noindent
3. Now let us consider the amplitude $A_5$ with  pion quantum numbers in
$t$-channel. This amplitude originates from the last ($\gamma_5$) term on
the r.h.s. of (\ref{3}). To avoid the admixture of the $t$-channel gluon state,
we choose here the isovector component \footnote{It is known that due to the
$\gamma_5$-anomaly in singlet channel, an admixture of the two-gluon state
alters strongly the mass and other properties of the $\eta^\prime$-meson.}.
In the Born approximation the high energy spin-flip amplitude is
\begin{equation}
\label{4}
A_5^B\left( \lambda=\frac 12, \lambda^\prime =-\frac 12;\,
-\frac 12,\, \frac 12\right) \,\propto \,\alpha_s\frac ts \, .
\end{equation}
At high orders of $\alpha_S$ the amplitude $A_5^B$ is multiplied by the same
DL function $F_{DL}^{(+)}$ (\ref{2}) for  even signature exchange (or by the
corresponding function $F_{DL}^{(-)}$ for odd signature exchange). Indeed let
us consider the first loop diagram Fig.2a.
With the help of Firz identity
(\ref{3}), we replace the $\gamma_\mu \cdot \gamma_\mu$ vertices of
the upper gluon in Fig.2a by the effective $\gamma_5$ vertex in Fig.2b
in order to select the contribution with  pion quantum numbers
in $t$-channel.

It is convenient to use the light-cone Sudakov variables
\begin{equation}
\label{5}
k_\mu\,=\,-\,\alpha\,p_{A\mu}+\,\beta p_{B\mu}+\,k_{\mu t}\,,
\quad d^4 k_\mu\,=\,\frac s2 \,d\alpha \,d\beta \,d^2k_t\,,
\quad s \simeq 2p_A\,p_B  \, ,
\end{equation}
and to close the contour of the $\alpha$-integral on the lower gluon
pole $1/(p_B-k)^2$. Then the loop integral reads\footnote{We consider here
the ladder type loop only as the non-ladder DL correction comes from the pure
classical current emission and does not depend on the spin structure
of the amplitude at all.}
 \begin{equation}
 \label{6}
 C_F\,\frac{g^2}{16\pi^3}\int \frac{d^2k_t}{k^4}\,
 \frac{s\, d\beta}{s\beta-k_t^2}\,\sum_{\mu=1,2}
 \overline u(p_B^\prime,\lambda^\prime) \gamma_\mu^\perp \widehat k \gamma_5
 \widehat k^\prime \gamma_\mu^\perp u(p_B,\lambda)
 \end{equation}
 $$
 \simeq
 C_F\frac{\alpha_S}{2\pi}\int \frac{d\beta}{\beta}\frac{dk_t^2}{k_t^2}\,
  \overline u \gamma_5 u,
 $$
where within the DL approximation we assume $k^2 \simeq k_t^2$ and
$k_t \gg q_t=(p_B-p_B^\prime)_t$; therefore $k_t^\prime \approx k_t$ and
$(k^\prime k) \simeq k^2$.
In addition we account for the fact that the
$s$-channel gluon $(p_B-k)$ is on-mass-shell and has only two transverse
polarizations $\mu =1,2$ (see \cite{10,6} for more details).

The interval of the $d\beta$
integration is restricted by the same DL constraint $\beta \gg k_t^2/s$
as that of the main (vector or axial in $t$-channel) component of the quark-quark
amplitude \cite{6,10}. Thus the first loop correction takes the form
\begin{equation}
\label{7}
\frac 12\,\frac{C_F\alpha_S}{2\pi}\,\ln^2\frac{s}{|\,t\,|}\,=\,\frac 12\,n \, ,
\end{equation}
which is equal exactly to the corresponding term in the expansion of
the DL function $F_{DL}$ (\ref{2}). The same is valid for the
DL contribution from higher loops. As  was shown in \cite{6},
each higher DL correction
does not depend on (or alter) the spin structure of the quark-quark
amplitude.

There is an analogous situation for the scalar  component
(the first term in (\ref{3})).
 Within the DL approximation the scalar and pseudoscalar exchange
are degenerate, just as are  the vector and axial components, discussed
above.

\noindent
4. Thus at small distances the quark-quark QCD interaction, mediated by the
$q \overline q$ pair ($t$-channel) exchange, contains the contributions
coming from the vector, axial, scalar and pseudoscalar terms in the Born
amplitude. Within the DL approximation these contributions never mix with
each other. The higher $\alpha_S$ order DL loop corrections simply multiply
each term of the Born amplitude by the same function $F_{DL}(\sqrt{\alpha_S}
\ln s/|t|)$; $F_{DL}^+$ (or $F_{DL}^-$) depending on the even (or odd)
signature of the amplitude. The 'vector' and 'axial' components conserve
the $s$-channel quark helicities, while the 'scalar' and 'pseudoscalar'
components flip helicities of the incoming quarks. However these spin-flip
amplitudes are suppressed at high energies by one power of $s$;
the pion-like, pseudoscalar exchange contains an extra factor $|t|/s$ in
comparision with the 'axial' contribution. Recall that the 'axial'
($A_1$-like) component describes the spin-spin correlation, that is the
difference of the non-flip amplitudes with different helicities of the
incoming quark.

{
Contrary to the "soft" Regge phenomenology (where the intercepts
$\alpha_{A_1}(0)\le 0$, $\alpha_\pi(0) \simeq 0$ of the $A_1$-
and $\pi$-trajectories
are smaller than intercepts $\alpha_R(0) \simeq 0.5$ of a natural
parity $\omega$, $\rho$,
$A_2$-trajectories)  the perturbative QCD DL
amplitudes with natural and unnatural parities are degenerate; the vector
($\gamma_\nu$) term is degenerate with the axial ($\gamma_\nu \gamma_5$)
one, and the scalar term with the pseudoscalar contribution. Therefore
at large scales (i.e. small distances) the high energy spin-spin correlation
occurs mainly due to the perturbative QCD interaction;
the non-perturbative $A_1$-exchange dies out with energy. On the other hand
the effective intercept,  $\omega_\pi$, of the perturbative QCD amplitude with
 pion $t$-channel quantum numbers $\omega_\pi = \omega_0-1$ is much
smaller than $\alpha_\pi(0) \simeq 0$. So, for a not too large scale, or
$|t|$, the spin-flip (unnatural parity) quark-quark interaction should be
described mainly by the non-perturbative pion Regge trajectory.
\sloppy

}
\vskip 1cm
{\bf Acknowledgements}\\
\smallskip
We are grateful to Prof. J. Bartels who had initiated this work. The work
was supported in part by the UK Particle Physics and Astronomy Research Council
and by the NATO Collaborate Linkage Grant SA (PST.CLG.976453)5437.
\vskip 0.9cm
\centerline{Figure captions}

\noindent
Fig.1 (a) The quark-quark,   and, (b) quark-antiquark Born amplitudes. Higher loop
corrections are shown in Fig.1(b) by dotted lines.\\
Fig2. (a) The first loop $q\bar q$-amplitude  and, (b) the pseudoscalar (
$t$-channel) component of the amplitude .

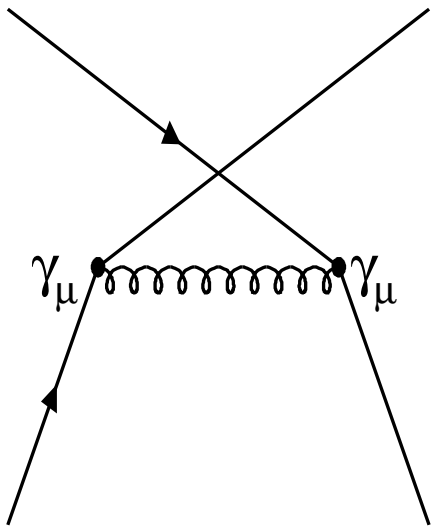
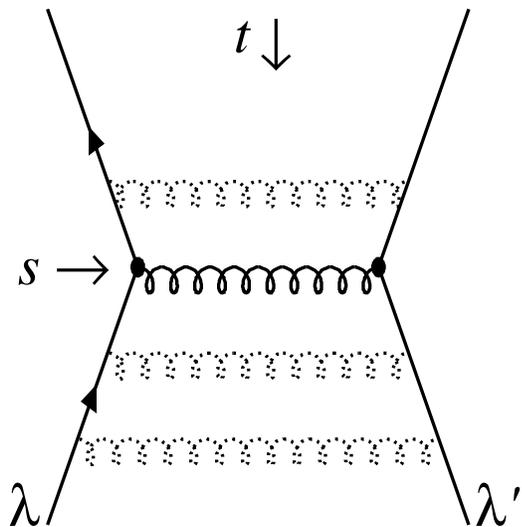
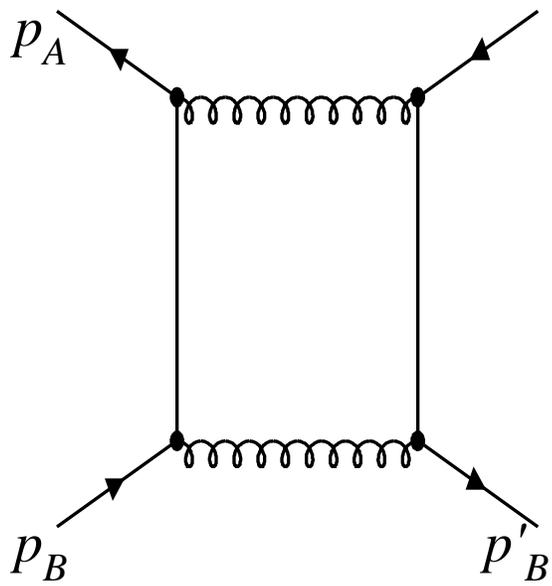
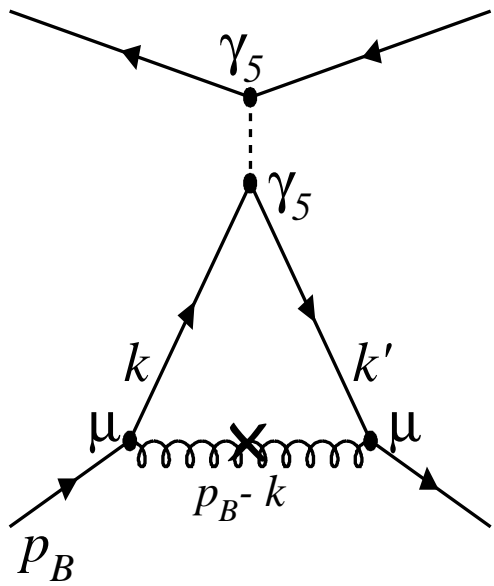
\end{document}